\documentclass[11pt,letterpaper]{article}

\usepackage[margin=1in]{geometry}
\usepackage[T1]{fontenc}
\usepackage[utf8]{inputenc}
\usepackage{graphicx}
\usepackage{booktabs}
\usepackage{subfig}
\usepackage{amsmath}
\usepackage{bm}
\usepackage{hyperref}

\begin{document}

\title{Performance and optimization of multi-fin oscillatory propulsion}

\author{Parker T. Grobe, Ethan J. Watson, and Tyler Van Buren\\[0.5em]
\small Department of Mechanical Engineering, University of Delaware\\
\small Newark, Delaware 19716, USA}

\date{}

\maketitle

\begin{abstract} Multi-fin bioinspired propulsion systems have seen a rise in interest recently due to natural occurrence in animal systems and emerging technology. These systems make use of fin-fin interaction to improve swimming performance where the downstream fins use the wake of the upstream fin to their benefit. While 2-fin systems are relatively well understood, it is unclear how performance is impacted as the number of fins grow. Using an immersed boundary method solver, we model the multi-fin systems to analyze performance across several key parameters: spacing between fins, phase offset, freestream velocity, and number of fins per system. We characterize performance via system thrust and propulsive efficiency. We show that a dimensionless phase parameter incorporating inter-fin spacing, oscillation frequency, and freestream velocity approximately collapses the thrust response across different operating conditions, providing a useful predictor of favorable wake-interaction timing. As more fins are added ($>2$), they can further enhance the group thrust and efficiency without modifying the performance of the fins upstream. Through optimization, we found this enhancement to peak with a 3-4 fin system, after which performance diminishes with additional fins due to the growing wake complexity. Optimal conditions led to boosts in thrust or efficiency of 30\% and 20\%, respectively. From this study, we show that there are performance gains realized in larger fin systems (though with diminishing returns) and optimizing system performance of multi-fin systems can be done sequentially to save computational resources.

\end{abstract}

\section{Introduction}
Bio-inspired propulsion systems have been steadily growing in popularity over the past few decades, particularly those utilizing fin-based oscillatory motions for subsurface vehicles \cite{ClarkSmits2006}. Such systems promise significant enhancements in thrust, stealth, and efficiency; crucial for the advancement of uncrewed underwater vehicles (UUVs) \cite{Tangorra2007, Fish2020, Zhou2024}. Foundational research has underscored the performance advantages of bio-inspired approaches \cite{Theodorsen1935, Lighthill1970}, with ongoing studies affirming the substantial potential in performance of these systems \cite{Smits2019}. Employing oscillating hydrofoils that execute heaving (lateral motion) and pitching (rotational motion) or their combinations enables the generation of predictable thrust and efficiency \cite{Floryan2017, VanBuren2019}.

Single fin systems have thus far been the primary focus of the community, analyzing the swimming performance of an oscillating propulsor. However, there are instances in both engineering and nature where oscillating propulsors interact. For example, the Plesiosaurus is thought to have used its multiple pairs of pectoral fins in unison to gain swimming benefit \cite{muscutt2017four}. Additionally, when fish school, they inherently interact with the wakes of the upstream swimmers---thought to lead to swimming performance gains and a potential motivating factor for schooling \cite{Saadat2021}. These inspirations are leading to new technologies that utilize fin-fin interaction, including the Boeing Wave Glider. Thus, we aim to explore multi-fin systems more generally with special consideration to systems where the fin number is $\geq3$. 

\begin{figure}[t]
  \centering
  \includegraphics[width=.9\linewidth]{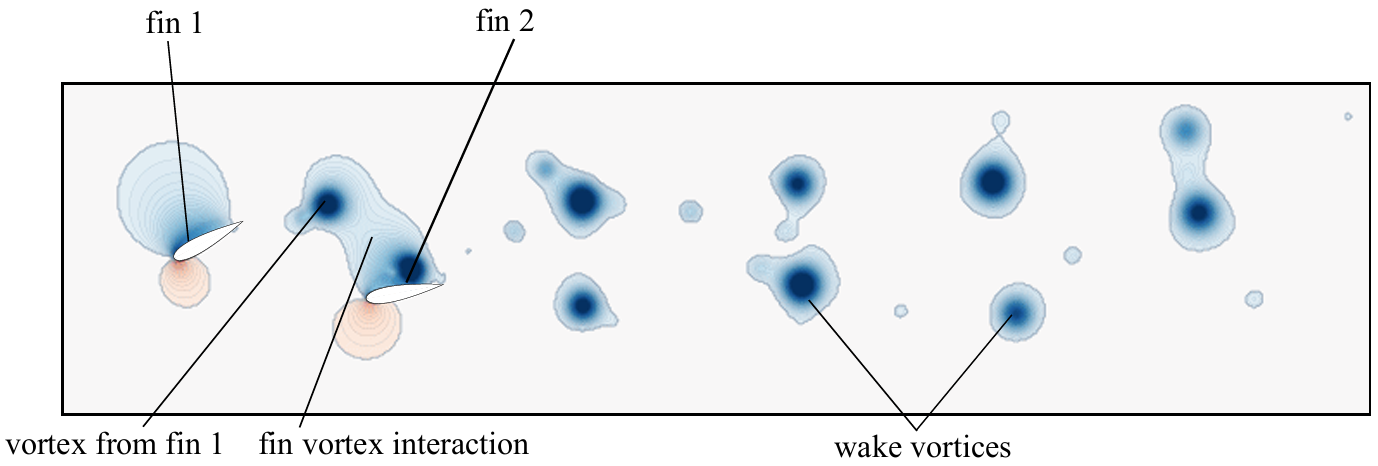}
  \caption{Sample pressure field of an inline 2-fin system wake interaction.} 
    \label{fig:flow_field_explanation}
\end{figure}

Transitioning from a single fin system to a two in-line fin system introduces increased complexity but with marked potential benefits. Research suggests that adding a second fin can range from doubling the thrust compared to a single fin or completely negate thrust production \cite{Muscutt2017}, depending on timing. The performance of such 2-fin systems hinges critically on the precise adjustment of inter-fin parameters, influenced by the lead fin's vortex shedding in the flow. This is depicted in the pressure flow field presented in figure \ref{fig:flow_field_explanation}, where regions of low pressure concentration represent vortex structure. The vortex structure is shed into the freestream and encountered by the second fin. For improved performance, the second propulsor needs to capture the vortex at the right time to enhance performance---much like a batter needs to time their swing in baseball or cricket. 
Extensive studies have shown that optimal performance is achieved when the downstream fin passes near to but not into the vortex shed by the upstream fin \cite{Muscutt2017}. This proximity allows the low-pressure region from the shed vortex to enhance the lift-based thrust at the leading edge of the second fin, thus boosting the overall system's thrust and efficiency. An additional consideration is the rotational velocity field advected by the vortex will directly influence the effective angle of attack of the downstream fin, leading to performance alteration, as described in \cite{SeoMittal2022}. We see this phenomenon occur in figure \ref{fig:flow_field_explanation} with the second fin's leading edge connecting and intensifying with the vortex shed by the lead fin. 

The progression from two to three or greater in-line fin systems in this framework has been less studied, and raises new questions about the scalability of fin-fin interactions. Here we explore the effects of adding additional fins to multi-fin system, investigating if the ideal parameters for 2-fin systems are applicable to systems with three or more fins. Furthermore, we seek to determine the ideal parameters for maximizing the efficiency and thrust of multi-fin systems and to assess whether adding more fins results in incremental benefits or diminishing returns. Insights from this study may influence future UUV designs as well as help us better understand natural swimmers. Note that, in this work each fin is directly downstream of another without a lateral offset and the spacing between each fin is steady while swimming. Thus, this work directly applies to systems where one central body controls the fins---as in the Plesiosaurus and Boeing Wave Glider. However, in fish schooling, keep in mind there is a lot more freedom in the propulsors to move or rearrange that should be considered \cite{Saadat2021, Liu2017, Guo2023, Weihs1973}. 

\section{Methods}
We conduct a simulation of oscillating foils in a fluid under a constant freestream velocity. The foils are a NACA 0020 shape with a chord of $c$ and have both sinusoidal heaving ($h$) and pitching motion ($\theta$) defined as:
\begin{equation}
    h = h_0\,\sin(2\pi ft + \phi)
\end{equation}
\begin{equation}
    \theta = \theta_0\,\sin(2\pi ft + \phi+\frac{3\pi}{2})
\end{equation}
where $h_0 = c$ and $\theta_0 = 30^\circ$ are the motion amplitudes, $f$ is the oscillation frequency, and $t$ is time. The pitching motion has a constant phase offset relative to the heave that produces near the best performance based on past research \cite{VanBuren2019}. The kinematics are generally defined in these systems via the Strouhal number $St=fh_0/U_\infty$ (heave motion dominates the wake width) which for this work is largely $St = 0.32$. Past work on multi-fin systems finds that performance changes due to fin interactions are largely consistent across a wide range of Strouhal numbers \cite{Muscutt2017}. The Reynolds number throughout is $Re = U_\infty \,c/\nu = 5000$.

Swimming performance is quantified through thrust and efficiency, which can be thought of as conduits for speed and cost of transport for a swimming system. The side forces ($F_y$), which act perpendicular to the freestream velocity, the moments ($M_z$) about the $z$-axis, the heave velocity ($\dot{h}$), describing the vertical oscillatory motion of the foil, and the pitch velocity ($\dot{\theta}$), capturing the rate of change of the foil's angular position, allow us to calculate the coefficient of power ($C_P$) as follows: 
\begin{equation}
    C_P = \frac{F_y \dot{h} + M_z \dot{\theta}}{\frac{1}{2} \rho U_\infty^3 s c},
\end{equation}
where $\rho$ is the fluid density, $U_\infty$ is the freestream velocity of the fluid, $s$ is the span of the foil, and $c$ is the chord length of the foil. Paired with measuring the thrust output $F_x$, which is the force acting along the direction of the freestream velocity, we can calculate the coefficient of thrust ($C_T$) as:
\begin{equation}
    C_T = \frac{F_x}{\frac{1}{2} \rho U_\infty^2 s c}.
\end{equation}
These coefficient definitions are traditional to hydrodynamics and have some momentum within the community as being the standard. However, we do want to recognize that perhaps the more appropriate velocity scale for non-dimensionalization should be based upon the lateral velocity of the foil, which has been used in studies \cite{van2018flow, yudin2023propulsive, wen2013understanding,bale2014energy,hieber2008immersed,zhou2012simulation}. Together, these coefficients allow us to determine the Froude efficiency ($\eta$) of the overall system, which is defined as:
\begin{equation}
    \eta = \frac{C_T}{C_P}.
\end{equation}

In this study we used an open-source computational fluid dynamics solver called WaterLily \cite{Weymouth2024}. WaterLily is an immersed boundary method framework that solves the incompressible Navier-Stokes equations on a uniform Cartesian grid using a finite-volume formulation. In its core implementation, WaterLily does not employ a separate turbulence closure model; instead, it operates as an implicit large-eddy simulation (iLES) solver, in which the dominant resolved vortical structures are captured directly while the effects of unresolved smaller scales are regularized through the numerical discretization. The solver has successfully been demonstrated to simulate a wide range of flows over complex 2D and 3D shapes such as cylinders in oscillating flows, a three-dimensional Taylor-Green vortex flow, and an oscillating propulsor system previously mentioned \cite{Weymouth2024, Muscutt2017}. 

To ensure that our simulations were adequately resolved, we conducted a detailed refinement study considering kinematic convergence, temporal resolution, and cell resolution. Full details and results from this convergence studied can be found in the Appendix. The kinematic study evaluated the number of cycles required for convergence, where a cycle is defined as the foil heaving through its full amplitude (two chord lengths) and returning to its initial position. Temporal resolution refers to the number of data points recorded per cycle, while cell resolution is defined by the number of grid cells per chord length. To balance accuracy with computational cost when exploring large parameter spaces, we adopted a two-stage approach: coarse-resolution simulations were first used to construct initial performance maps and identify regions of interest in multi-fin systems, after which regions of interest were recomputed at higher resolution to obtain final performance metrics and enable detailed analysis of force production and flow-field dynamics. For the coarse search, performance metrics were computed using the third oscillation cycle with 33 data points per cycle and a spatial resolution of 32 cells per chord. For the fine-resolution simulations, three oscillation cycles were used at minimum with an additional cycle for each added fin beyond $N=3$, but with 100 data points per cycle and a spatial resolution of 64 cells per chord. (Select cases were also run at 128 cells per chord to confirm trends obtained were valid at lower resolutions). Given that the present framework is two-dimensional and uses an iLES formulation without an explicit turbulence model, the refinement studies are intended to assess convergence of the resolved large-scale wake structures and integrated force metrics, rather than to claim full resolution of the smallest vortical scales in the most disordered wakes. The results of the convergence studies and further validation explanations are presented in Appendix A.

\section{Results and Discussion}
\label{sec:ResultsAndDiscussion}
With this work, we strive to better understand the performance of in-line oscillating propulsor systems with a focus on the system thrust and efficiency as performance metrics. We will begin with a review of a two fin system, then move into a parametric exploration of the three fin system, finally moving to optimization of systems with up to six fins---ideally finding some generalized asymptotic behavior that may apply to $N$-fin systems. Variables of interest will be fin spacing, phase between fin oscillations, and freestream velocity. 

\subsection{Two-fin system}\label{sec:2fin}
\begin{figure}
    \centering
    \includegraphics[width=1\linewidth]{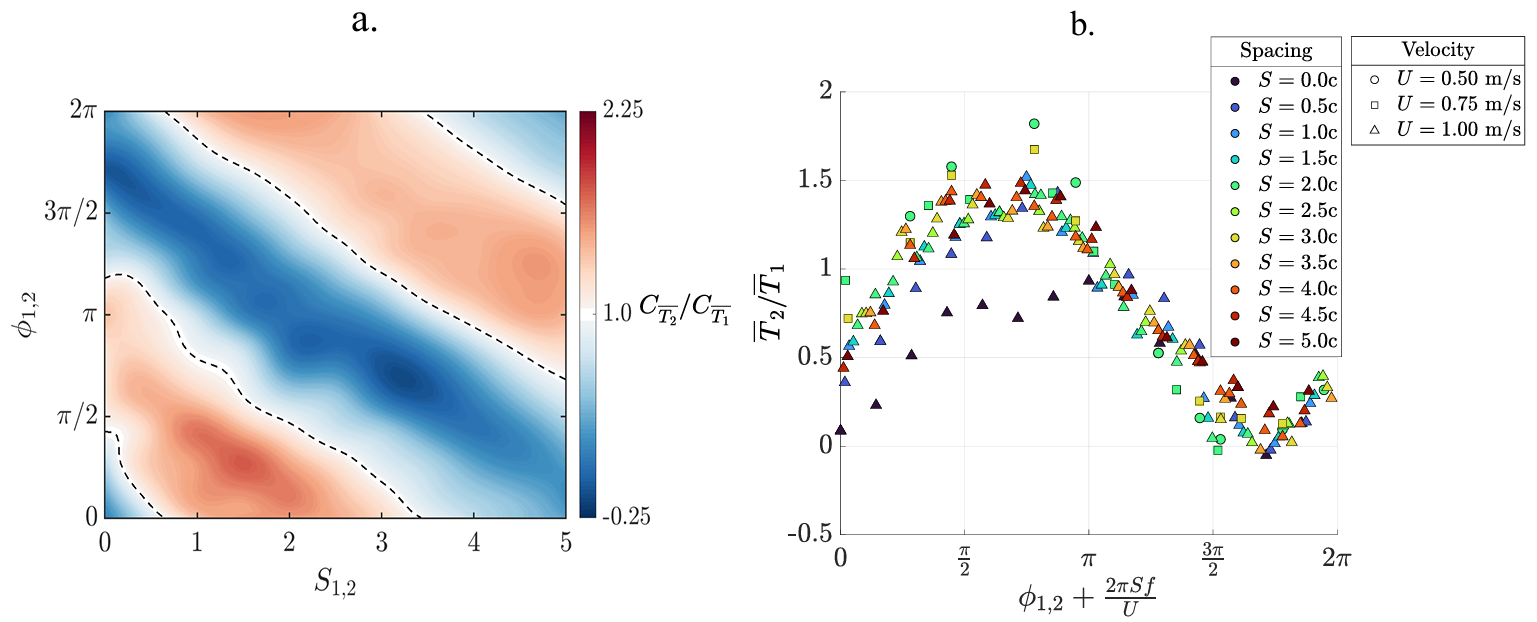}
    \caption{Thrust of the downstream fin relative to the upstream fin in a 2-fin system as it varies with phase offset \& spacing (a) and compared to a phase offset with a phase modifier parameter to account for wake timing (b).}
    \label{fig:2fin_proof}
\end{figure}

As a community, we have a good understanding of the performance benefit of in-line 2-fin systems thanks to work like \cite{Muscutt2017}. For both validation and further exploration,  in figure \ref{fig:2fin_proof}a we recreate these results where both fin spacing $S$ and phase $\phi$ are systematically adjusted. A salient feature of this performance space are the linear sloped thrust contours. This means that if the fin spacing were to change, the phase would need to proportionally change to maintain performance. This is logical in that there is an ideal timing to which the downstream fin hits a vortex shed from the upstream fin---a timing that is impacted by spacing, phase, as well as freestream velocity. This non-dimensional timing can be simply represented as the proportion of the time-delay, based on the vortex propagation velocity (approximated as the freestream velocity for a reference scale) and distance between fins $\sim S/U_\infty$, with respect to the period of fin oscillation, $1/f$. This gives that the ideal phase modifier between the fins is

\begin{equation}
    \phi_0 \sim 2\pi\frac{S f}{U_\infty}.
    \label{eqn:phaseMod}
\end{equation}

This idea of an ideal phase modifier describing the timing conditions required between fins for optimal vortex interaction has been explored previously in research on energy harvesting systems utilizing fins \cite{Dumas2012, XU2017202, ribeiro2021wake}. To show this relationship, we plot the thrust coefficient as it varies with phase offset of the second fin ($\phi_{1,2}$) combined with the phase modifier ($\phi_0$) in figure \ref{fig:2fin_proof}b. For a wide range of fin spacings, phases, and freestream velocities, the values roughly collapse. With the phase modifier accounted for, the ideal phase of the second fin relative to the first is about $\phi_{1,2} = 2\pi/3$. Note here that these results span Strouhal numbers of $St = 0.16-0.32$.

\subsection{Three-fin system}\label{sec:3fin}

\begin{figure}
  \centering
  \includegraphics[width=.9\linewidth]{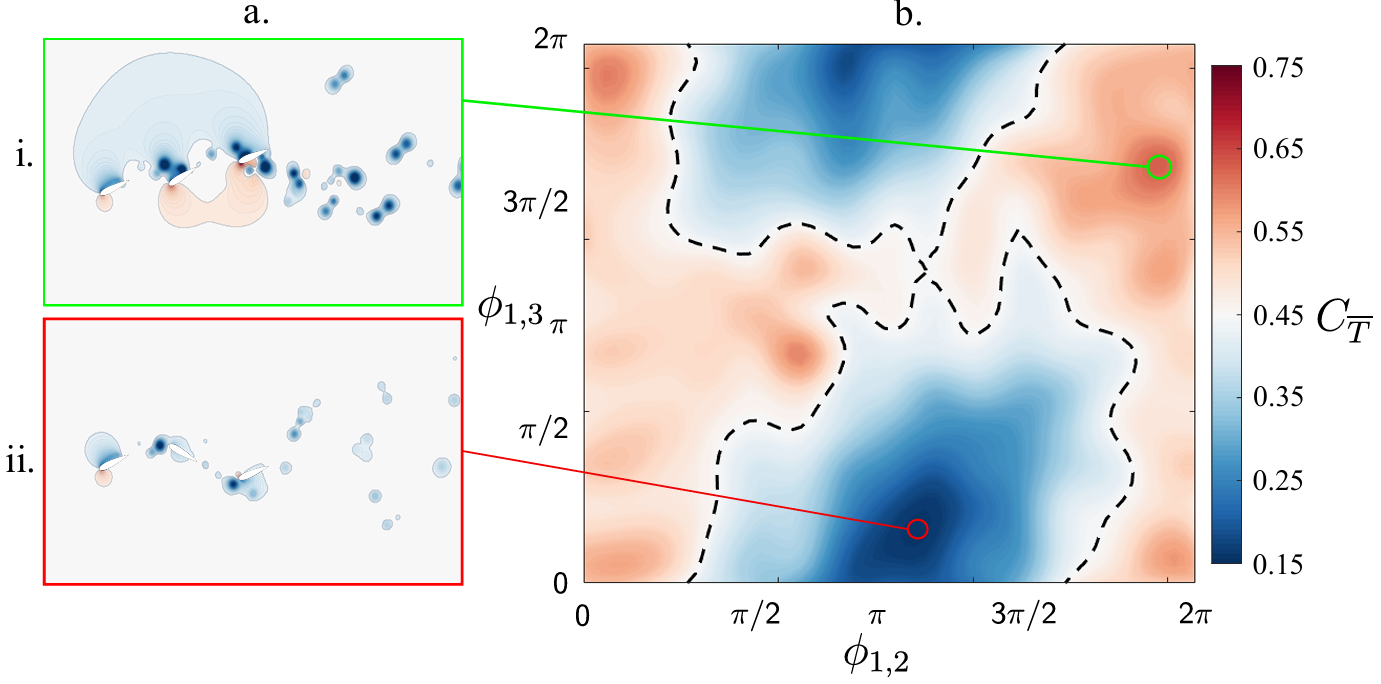}
  \caption{Thrust of a 3-fin system as it changes with the phases between fins (b) for a fin spacing of $S=2$ with a representative flow field at the best and worst performance conditions (a). Supplemental files have animations of these same flow fields \cite{SM}.}
  \label{fig:3_fin_thrust_map}
\end{figure}

We next explore the basic 3-fin system parametrically with variable fin spacings and phases. Here, we will restrict the fins to have equal spacing (unequal spacing is considered later in our optimization). Figure \ref{fig:3_fin_thrust_map} shows a complete performance map of the system mean thrust coefficient $C_{\overline{T}}$, spanning 225 total cases, as it varies with the phase offset of fin 2 relative to fin 1, $\phi_{1,2}$, and the phase offset of fin 3 relative to fin 1, $\phi_{1,3}$. For all the cases, the spacing between fins is held at two chord lengths (foil spacing is considered later in this work).
 
To the left of the performance map are two highlighted cases, showing the global pressure field for the best and worst case in the parameter space. Generally we see that the phase of the fins can lead to a dramatic change in system thrust, ranging from approximately $C_{\overline{T}} = 0.15 - 0.60$.  The distribution of thrust takes on an interesting structure, with vertical banding of high and low thrust indicating a periodic dependence on the phase between fin 1 and 2. However, one 45$^\circ$ band of higher thrust crosses the phase space, which implies that there is an ideal phase for fin 3 but if fin 2 has a phase change, then fin 3 must change similarly. While snapshots of an intrinsically dynamic flow field are limiting, we see that the flow for the best case has all of the low pressure regions from upstream vortex structures aligning with the low pressure region of the lift of the downstream foils. In the worst thrust case, the low pressure regions of vortex structure impede the higher pressure on the bottom side of the lifting foil, potentially canceling the lift effect that leads to thrust. For additional exploration and detail, we have added animations of each of these two cases to accompany figure \ref{fig:3_fin_thrust_map} as a supplemental file \cite{SM}.

\begin{figure}
  \centering
  \includegraphics[width=1\linewidth]{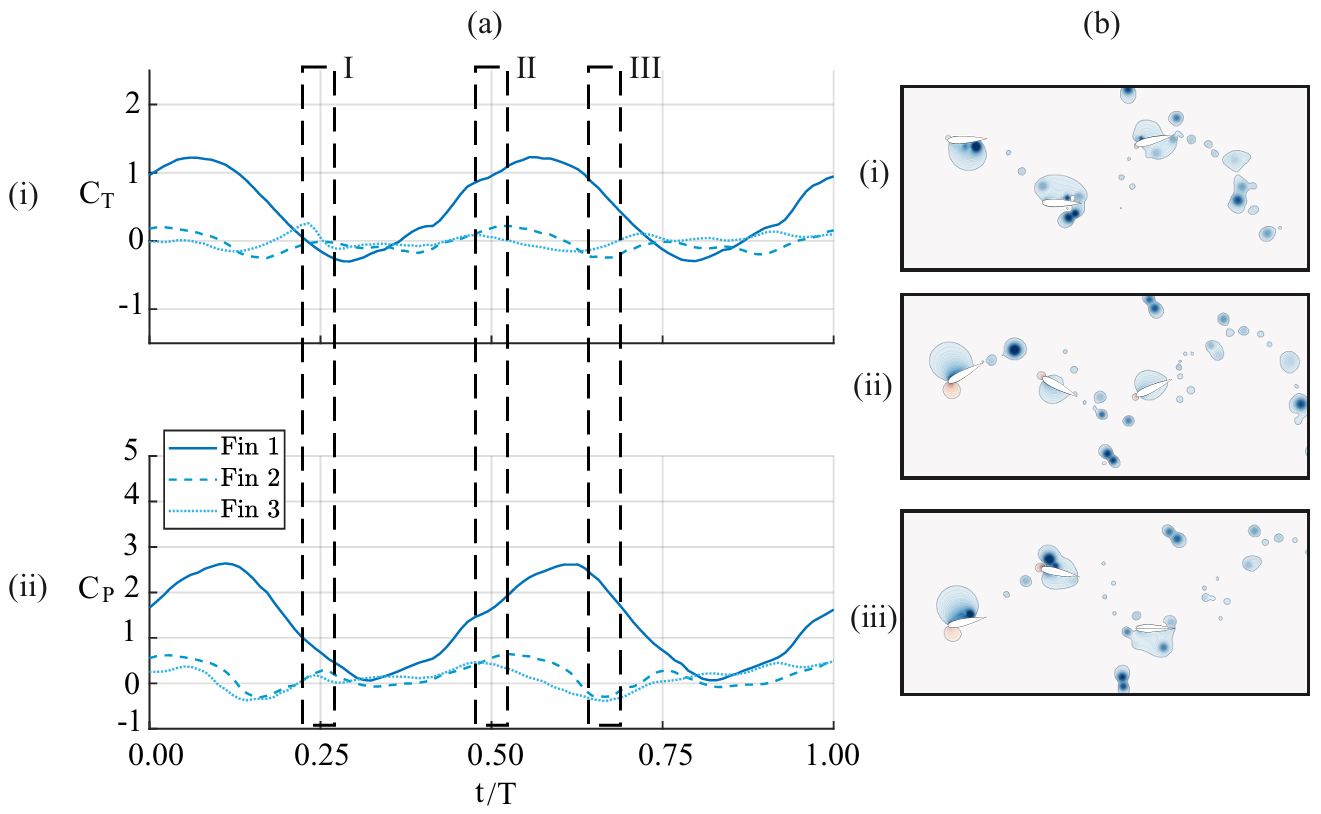}
  \caption{Instantaneous thrust and power coefficients over the final cycle for the low-thrust 3-fin system, paired with flow fields at selected times in the cycle. Panel (a) shows the force histories, and panel (b) shows the corresponding flow-field snapshots.}
  \label{fig:low_thrust_force_history}
\end{figure}

Figure \ref{fig:low_thrust_force_history} expands on the highlighted flow fields by showing the instantaneous thrust and power coefficients for the three fins in the low-performing case in panel (a). Three specific times in the cycle are marked on the force-history plots with dotted boxes and labeled I, II, and III. These marked times correspond to the flow-field snapshots in panel (b), with (b.i), (b.ii), and (b.iii) corresponding to I, II, and III, respectively. 

Looking at the three highlighted snapshots in the force history and flow-field animations, we may discern the potential causes for low performance, after which we will compare the high-performing case. Beginning with the first time capture, labeled I on the force-history plot in Fig.~\ref{fig:low_thrust_force_history}(a.i), we observe a low instantaneous coefficient of thrust from all three fins. In fact, fin 1 is beginning to enter a negative-thrust state, indicating that it is acting as a drag-producing body at that instant. This coincides with a low power requirement to drive the fin, as shown by the low coefficient of power in Fig.~\ref{fig:low_thrust_force_history}(a.ii). Comparing this moment with the flow-field snapshot in Fig.~\ref{fig:low_thrust_force_history}(b.i), all three fins are at their maximum heave positions in either the positive or negative lateral direction. This is consistent with the expectation that thrust production is low at the point where the fins reverse direction to begin the next half-stroke. Because the majority of the thrust arises from lift-based mechanisms, with added-mass effects contributing less overall thrust, this turnaround point corresponds to a minimum in thrust production. As the fins alter their lateral direction, the primary lift-based contribution has diminished, and the fins have begun to shed their vortices, indicating that the added-mass benefits have also largely concluded. 

Moving to the next highlighted moment, labeled II in Fig.~\ref{fig:low_thrust_force_history}(a.i), the thrust production of fin 1 rises well above that of fins 2 and 3. The coefficients of power in Fig.~\ref{fig:low_thrust_force_history}(a.ii) follow the same general trend as the thrust histories for each fin. Examining the corresponding flow snapshot in Fig.~\ref{fig:low_thrust_force_history}(b.ii), all three fins are in, or entering, the middle portion of their lateral heave motion. This suggests that peak thrust production occurs during the mid-stroke. At this stage, lift-based thrust is strongest, as the pressure differential across the two sides of each fin is near its maximum, with blue denoting low pressure and red denoting high pressure. In this snapshot, fin 1 exhibits a strong pressure differential, whereas fins 2 and 3 show a much weaker differential. Additionally, the downstream fins are moving through a more chaotic flow environment, which likely contributes to their reduced performance.

\begin{figure}
  \centering
  \includegraphics[width=1\linewidth]{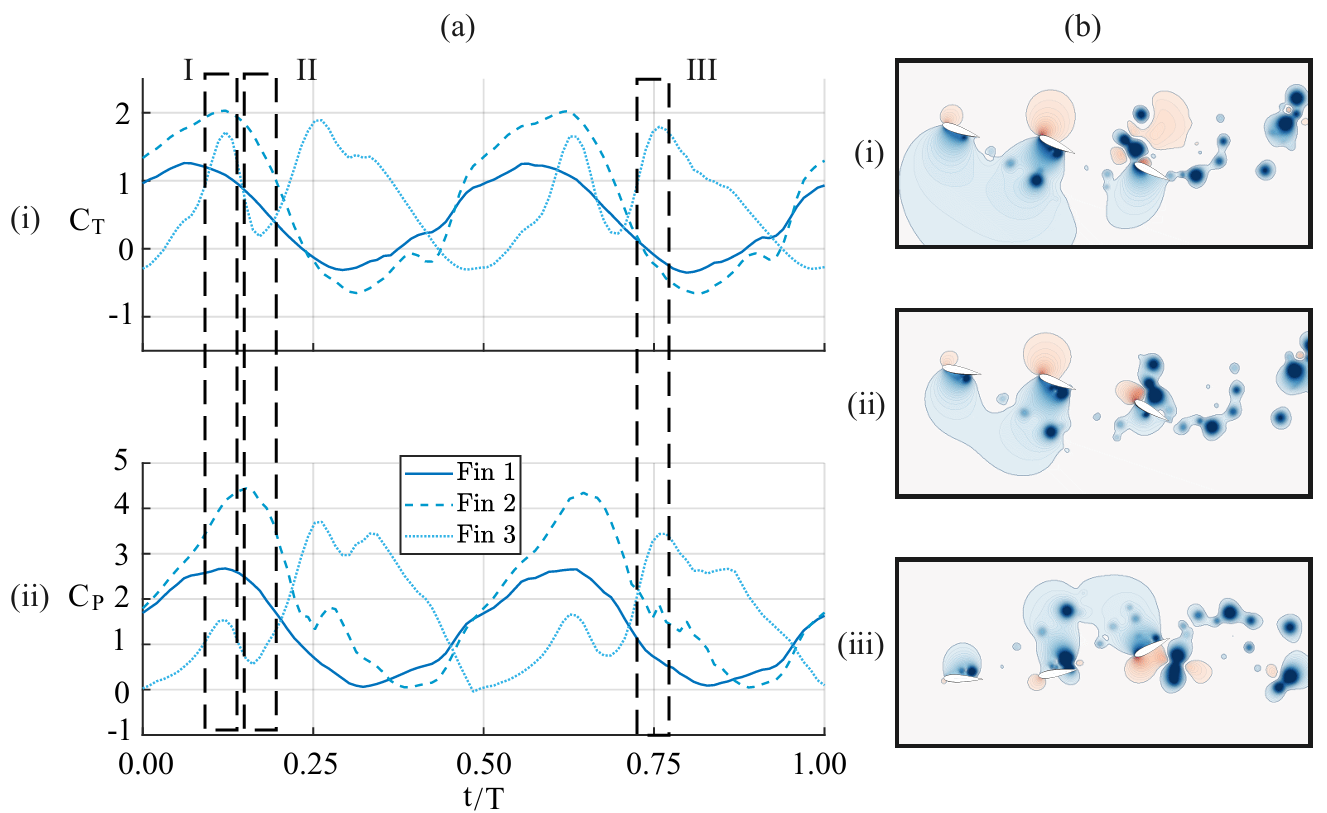}
  \caption{Instantaneous thrust and power coefficients over the final cycle for the high-thrust 3-fin system, paired with flow fields at selected times in the cycle. Panel (a) shows the force histories, and panel (b) shows the corresponding flow-field snapshots.}
  \label{fig:high_thrust_force_history}
\end{figure}

Lastly, at III in Fig.~\ref{fig:low_thrust_force_history}(a.i), the lead fin's thrust begins to trail off, while fins 2 and 3 reach their lowest thrust-production levels. Both fins 2 and 3 exhibit negative thrust, meaning that they are acting as sources of drag on the overall system. Comparing the force histories with the snapshot in Fig.~\ref{fig:low_thrust_force_history}(b.iii), we begin with fin 1. As it enters its lateral turnaround point, its thrust drops off, and the low-pressure region around the fin begins to roll up into the vortex that will be shed downstream. Fin 2, now acting as a drag producer, exhibits only a weak pressure differential. The same strong blue-red contrast across the leading and trailing surfaces is no longer present. Furthermore, comparing this with the earlier snapshot in Fig.~\ref{fig:low_thrust_force_history}(b.ii), a vortex shed from fin 1 can be seen convecting downstream toward fin 2. By Fig.~\ref{fig:low_thrust_force_history}(b.iii), that vortex has collided with the upper portion of fin 2, increasing low pressure on the trailing side of the fin. This reduces the lift-based thrust, disrupts the vortex-shedding process, and contributes to drag production. 

Matching the structure of Fig.~\ref{fig:low_thrust_force_history}, we plot the force histories and corresponding highlighted flow snapshots for the high-thrust case in Fig.~\ref{fig:high_thrust_force_history}. Regardless of the downstream fin phases, the lead fin displays the same general force behavior; thus, any difference in system performance lies in the behavior of the second and third fins. Starting with I in Fig.~\ref{fig:high_thrust_force_history}(a.i), fins 2 and 3 exhibit much higher thrust production than fin 1, unlike the low-thrust case shown in Fig.~\ref{fig:low_thrust_force_history}. This is accompanied by a much higher power requirement, as shown by the spike in coefficient of power in Fig.~\ref{fig:high_thrust_force_history}(a.ii). Fins 2 and 3 show a much stronger low-pressure region on their leading surfaces; the low pressure generated by their motion combines constructively with the low pressure associated with vortices in the freestream. Looking more closely at fin 3, however, vortices pass near the upper portion of the fin. Advancing slightly to the second highlighted moment, II, the third fin's thrust production drops dramatically. Viewing the flow in Fig.~\ref{fig:high_thrust_force_history}(b.ii), the low-pressure field associated with the vortices has disrupted the flow over fin 3, even causing high pressure to form near the leading edge, thereby producing a sharp reduction in forward thrust. Interestingly, this coincides with a drop in coefficient of power. In contrast, fins 1 and 2 continue through their heave motion, with their low-pressure regions remaining coherent and without strong negative interactions with nearby vortices.

Finally, at III in Fig.~\ref{fig:high_thrust_force_history}(a.i), fin 3 has recovered from its negative interaction with the shed vortices and reaches a peak in both thrust production and power usage. In the corresponding flow field, Fig.~\ref{fig:high_thrust_force_history}(b.iii), the pair of vortices that had previously disturbed the fin has now convected downstream past it. In addition, the low-pressure region on the leading surface of fin 3 interacts constructively with that of fin 2, while a strong high-pressure region develops on its trailing surface. Together, these features produce a strong favorable pressure differential and enhanced thrust. 

From these observations, several conclusions can be drawn. High thrust performance results from positive vortex interaction, meaning that vortices shed upstream pass near the leading surface of a downstream fin, where the low-pressure regions reinforce one another and increase the overall pressure differential. In contrast, low thrust performance arises when vortices collide directly with downstream fins or pass near their trailing surfaces, thereby reducing the pressure differential. In either case, both lift-based thrust and added-mass contributions are diminished, disrupting favorable force generation. Finally, the results also show that high thrust production is accompanied by high power consumption. 

Generally, these results imply that the ideal thrust-producing heuristics for a 3-fin system closely mirror those observed in 2-fin configurations as previously outlined \cite{Muscutt2017}. To confirm this, we need to explore the interaction between upstream and downstream fins more deeply. Specifically, whether upstream fins influence downstream fins solely via shed vortices, or if downstream fins also affect the upstream ones, and at what inter fin spacing influence becomes negligible. 

\begin{figure}
  \centering
  \includegraphics[width=1
  \linewidth]{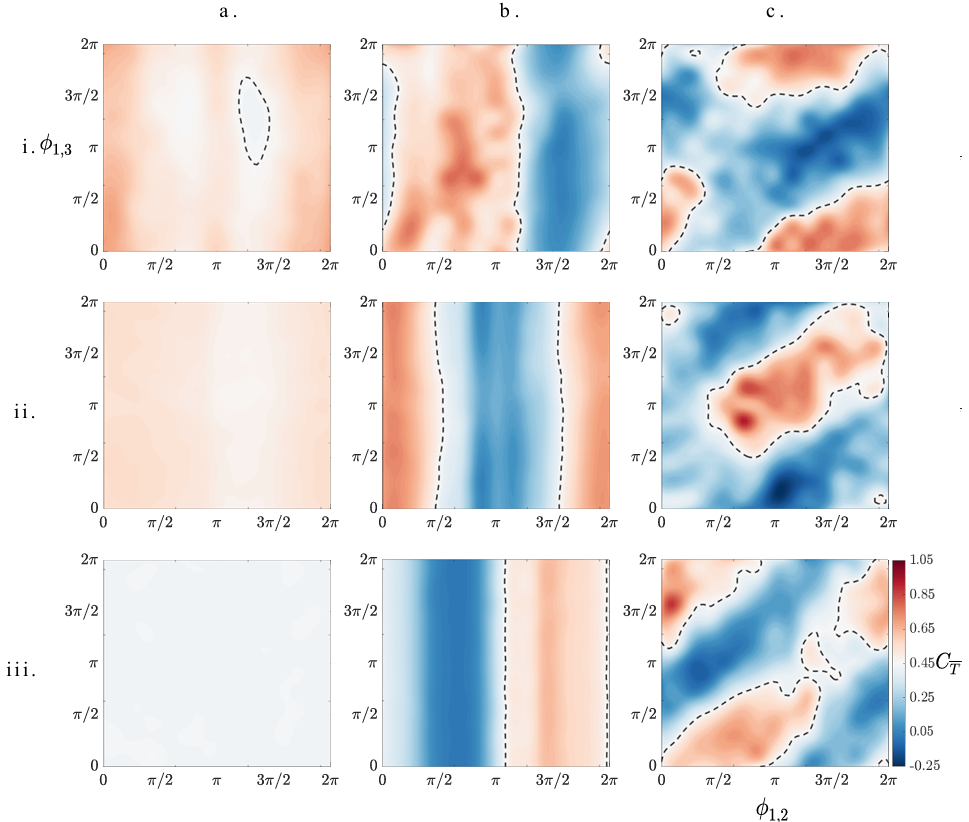}
  \caption{Thrust of the first, second, and third fins in a 3-fin system (a-c) as it changes with the phases between fins for a fin spacing of $S=0.5, 2, 10$ (i-iii).}
  \label{fig:3fin_thrust_varied_spacing}
\end{figure}

\begin{figure}
  \centering
  \includegraphics[width=1.0\linewidth]{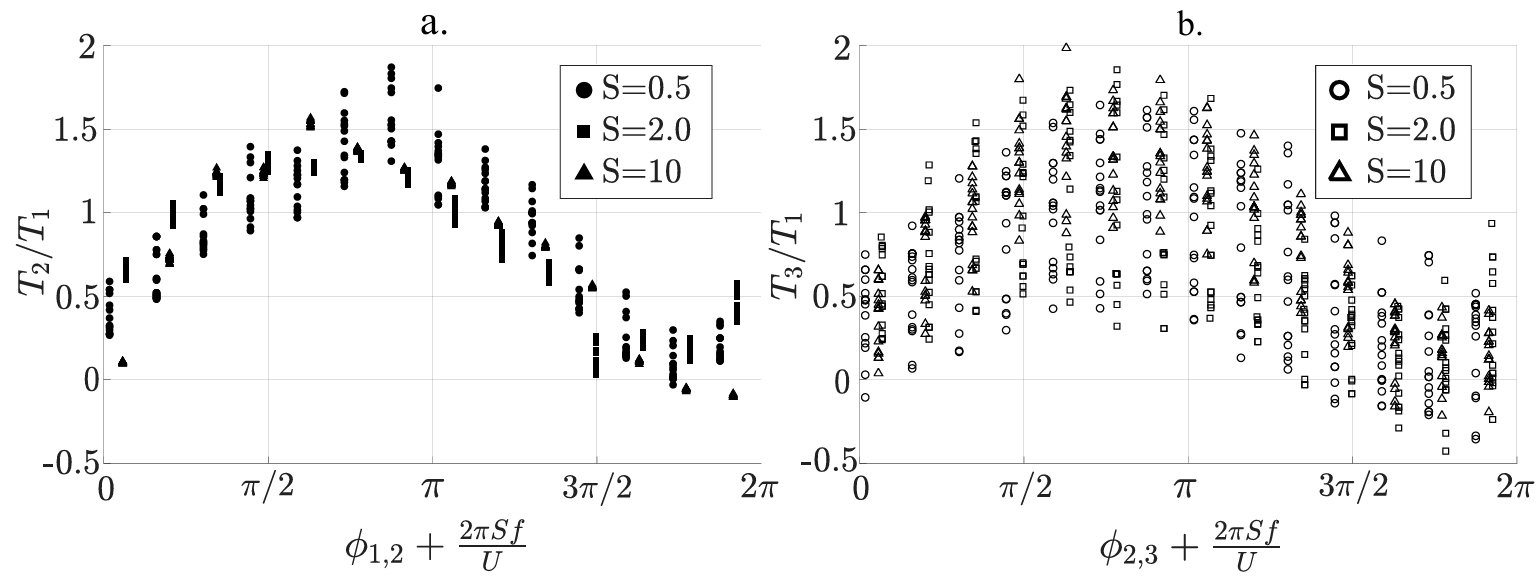}
  \caption{Thrust of the second (a) and third (b) fins relative to the lead fin as it varies with modified phase for the cases in figure \ref{fig:3fin_thrust_varied_spacing}.}
  \label{fig:phase_modifier_3fin}
\end{figure}

Consider a 3-fin system, but with variable spacing between fins. By bringing fins closer together, any impact on an upstream fin from a downstream one should be exaggerated. Figure \ref{fig:3fin_thrust_varied_spacing} shows the thrust as it varies with fin phases for fin spacings of $S=$0.5, 2, and 10 chords. (Note that, here we explore the mean thrust of individual fins $C_{\overline{T}_n}$.) The most upstream fin is relatively unaffected by a change in phase of the fins downstream with constant thrust across the parameter space, though the closest fin spacing shows minor periodic influence due to the nearest downstream fin. Even when fins are quite close together, we see only minor upstream effects. The second fin exhibits vertically banded periodic thrust, meaning there is an ideal phase $\phi_{1,2}$ for thrust production that is relatively uninfluenced by the downstream fin phase $\phi_{1,3}$. Finally, for the third fin, the thrust contours exhibit approximate diagonal banding with a slope near \(45^\circ\), though the pattern is less uniform than the cleaner bands observed elsewhere. This suggests that the third fin performance varies with both its own phase, \(\phi_{1,3}\), and the phase of the second fin, \(\phi_{1,2}\), indicating that its timing is influenced by the combined upstream wake environment rather than by the nearest upstream fin alone. In other words, while the second fin may provide the strongest direct influence on the third fin, the third fin performance is still shaped by the interaction of structures shed by both upstream foils. 

Now we see more concretely that for a three fin system, the upstream fins are generally unaffected by downstream fins unless very closely spaced. This has major implications for optimization of fin systems, as we will see later on. Consider a scenario where you have an optimized 3-fin system and choose to add another fin, you would only need to optimize the phase of the new fin while holding the others constant---greatly reducing the parameter space. We refer to this strategy later as \emph{sequential optimization}.

Using the data in the cases from figure \ref{fig:3fin_thrust_varied_spacing}, we can test the validity of the phase modifier $\phi_0$ formulated earlier on this 3-fin system. Figure \ref{fig:phase_modifier_3fin} shows the thrust for the second and third fin relative to the first fin across a range of phases for all three spacings. As with the 2-fin system, the phase modifier successfully aligns all the cases to a single periodic curve---again peaking around $\phi_{n-1,n} = 2\pi/3$. However, for the third fin the data is not nearly as collapsed and has a much wider performance band. This is because of the increased chaos in the wake (e.g., the wake of the first fin on the third) leading to other variables impacting performance not accounted for by the phase modifier. This effect is amplified as the fin spacing is smaller. However, the phase modifier is successful in capturing the performance trends and may be a useful tool for predicting the ideal fin timing in multi-fin systems.

\begin{figure}
  \centering
  \includegraphics[width=.9\linewidth]{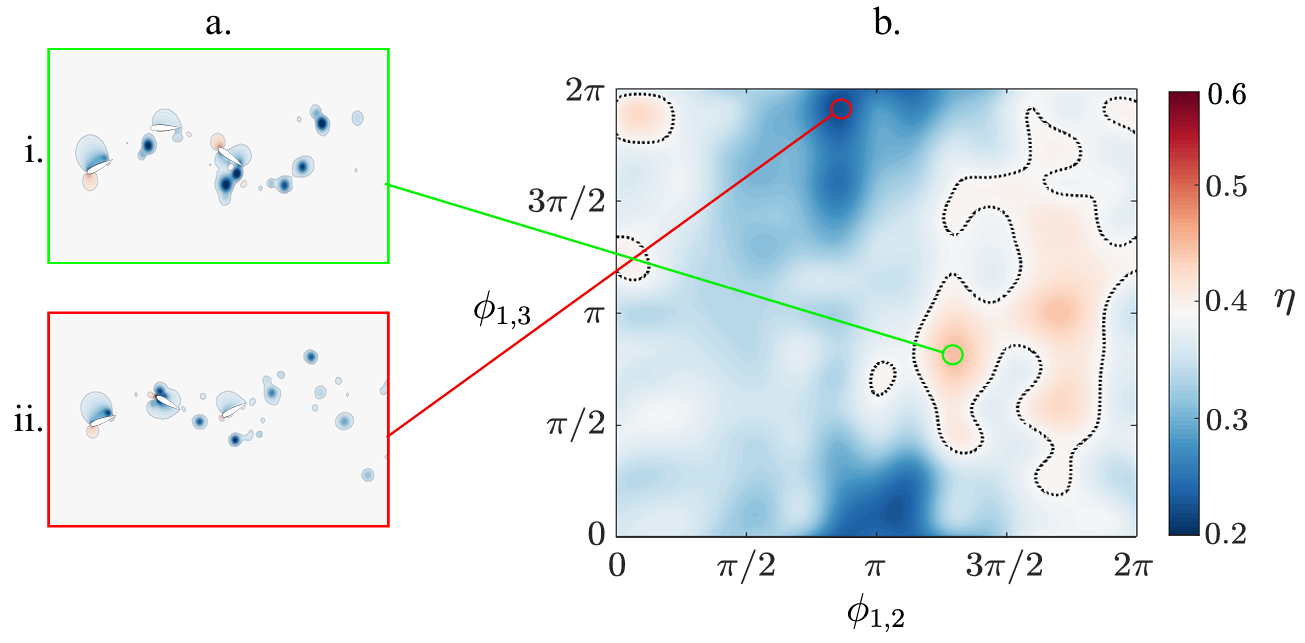}
  \caption{Efficiency of a 3-fin system as it changes with the phases between fins (b) for a fin spacing of $S=2$ with a representative flow field at the best and worst performance conditions (a). Supplemental files have animations of these same flow fields \cite{SM}.}
  \label{fig:3_fin_eff_map}
\end{figure}

To this point, we have considered only the thrust which is directly related to swimming speed. In comparison, we can explore the efficiency $\eta$ for the same parameters to see the impact multi-fin systems have on energetics. Previous studies \cite{Floryan2018} show that efficiency and thrust are generally inversely correlated for individual propulsors, but it is unclear if this holds true for multi-fin systems. Figure \ref{fig:3_fin_eff_map} shows the efficiency as it varies with the phase of the second and third fins, $\phi_{1,2}$ and $\phi_{1,3}$, respectively. Across the phase space, we see a range of efficiencies from 0.22 to 0.45, with a distribution that closely matches the thrust distribution shown back in figure \ref{fig:3_fin_thrust_map}. Regions of lower efficiency occur at phases of $\phi_{1,2} \approx \pi$ and $ -\pi/2 < \phi_{1,3} < \pi/2$, which aligns directly with regions of \emph{lower} thrust. However, regions of higher efficiency than average are less pronounced, and do not necessarily present themselves as clearly as high thrust. This means that, you can operate at regions of high thrust without an efficiency penalty, and if you operate at a region of low thrust you are also less efficient. 

For each region of low efficiency and high efficiency, we present a snapshot of the flow field colored with regions of pressure due to the vortex structure. In the lower efficiency region, which aligns with a region of poor thrust behavior shown earlier, we see that the low pressure regions are operating further from the foils and on the higher pressure side. For the high performance case, we see low pressure vortex regions interacting with the suction side of the foil, assisting in lift generation and thrust performance. In this context, the low pressure region is providing lift assistance without the local foil needing to generate the circulation---leading to higher efficiency.

\begin{figure}
  \centering
  \includegraphics[width=1\linewidth]{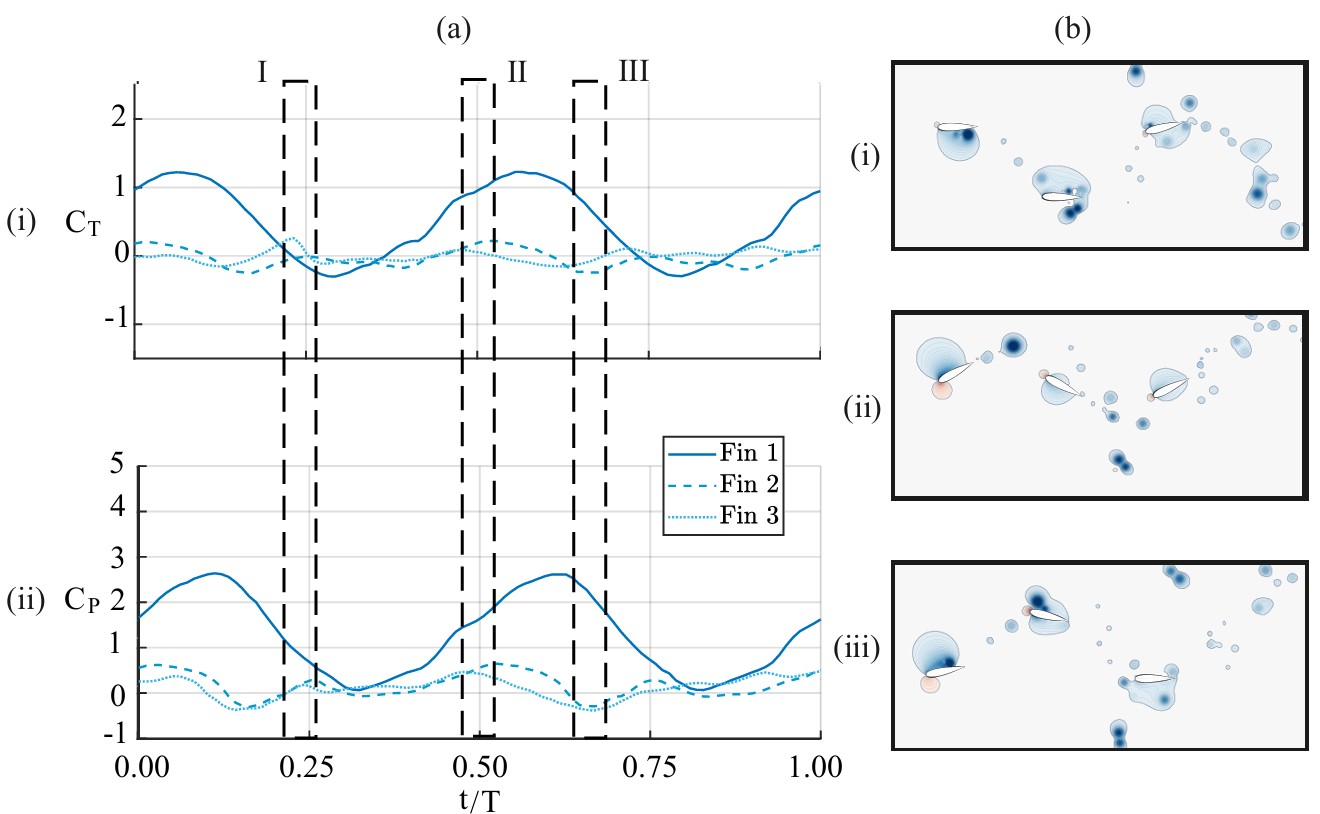}
  \caption{Instantaneous thrust and power coefficients over the final cycle for the low-efficiency 3-fin system, paired with flow fields at selected times in the cycle. Panel (a) shows the force histories, and panel (b) shows the corresponding flow-field snapshots.}
  \label{fig:3_fin_low_eff_analysis}
\end{figure}

\begin{figure}
  \centering
  \includegraphics[width=1\linewidth]{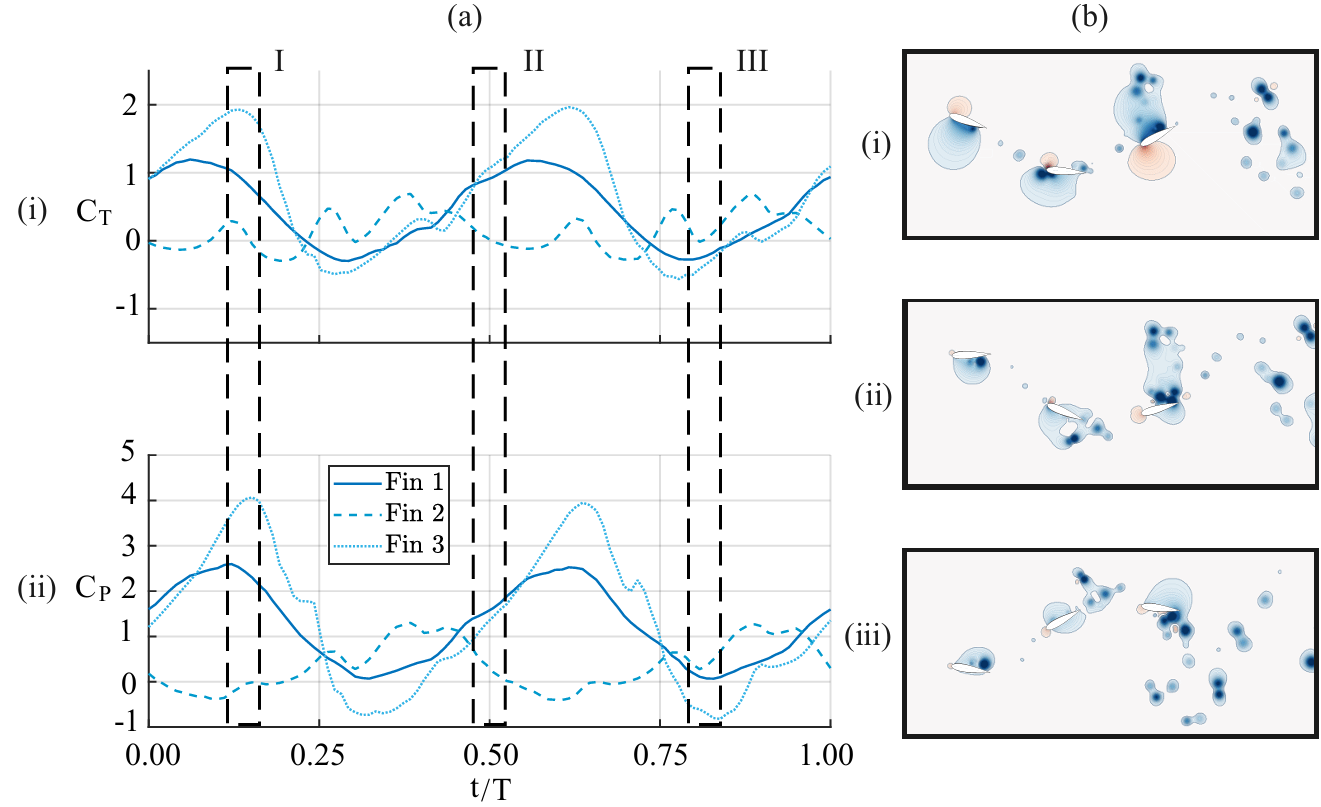}
  \caption{Instantaneous thrust and power coefficients over the final cycle for the high-efficiency 3-fin system, paired with flow fields at selected times in the cycle. Panel (a) shows the force histories, and panel (b) shows the corresponding flow-field snapshots.}
  \label{fig:3_fin_eff_analysis}
\end{figure}

To better understand this flow field timing and impact on efficiency, we can look to time-series of a period of the thrust and power coefficient for the three fins in the low and high efficiency regions, as shown in figures \ref{fig:3_fin_low_eff_analysis} and \ref{fig:3_fin_eff_analysis}. Each time-series has pressure fields at points of interest. As a reminder, efficiency is dictated by both thrust and power such that $\eta = C_T/C_P$. For the low efficiency region in figure \ref{fig:3_fin_low_eff_analysis}, we see similarity to the low thrust region discussed in figure \ref{fig:low_thrust_force_history}. (In fact, the low thrust and low efficiency regions are similar in phase offset combinations). The low efficiency stems from the second and third fins producing no thrust at all, often acting as drag sources. Where we do see interesting differences, however, is when we compare the high thrust region behavior to the high efficiency region. Recall, in the high thrust region of figure \ref{fig:high_thrust_force_history}, all three fins were large thrust producers---unsurprisingly. In that scenario, the second fin had the highest power draw of the three. In the high efficiency region shown in figure \ref{fig:3_fin_eff_analysis}, the first and third fin produce considerable thrust, but the second fin is in almost a zero thrust state throughout the cycle. In this configuration, the vortex interaction benefits seem to bypass a relatively neutral second fin, which leads to a considerable power savings for the system. These results suggest that optimal multi-fin propulsion does not necessarily require every fin to contribute equally to thrust production. Instead, the collective efficiency of the system can benefit from configurations in which individual fins assume different hydrodynamic roles. 

Overall, for a three fin system we see clear distinct variation of thrust and efficiency performance dictated by the phase of the downstream fins. For thrust, there are large swaths of phase regions that lead to higher performance and lower. However, for efficiency the higher performance region is more narrow and much more sensitive to changes in phase.

\subsection{Large fin system optimization}

If we want to consider more fins or variables in our system, the parameter space becomes too large to parse efficiently. We look to optimization to simultaneously find the best performing cases while also learning more about the limitations of multi-fin swimming systems for larger parameter spaces. For this work, we paired the immersed boundary method simulation with a Bayesian optimization. Bayesian based optimization possesses the ability to traverse a complex parameter space efficiently, making it an ideal optimizer for computationally expensive problems. 

In short, Bayesian optimization begins by defining the parameter space to the optimizer, essentially giving its bounds to explore within. Then it runs a given number of preliminary simulations to get a starting data point. Following this it will choose the next data point, as it collects more known or observed points it creates a best guess of the space between these points. This best guess is called a surrogate model, predicting a map of the parameter space between data points. The surrogate model is created via a Gaussian process, it defines a best estimate (mean function) between two known points and then an uncertainty range of possible other guesses around that mean. Between two points, the process predicts multiple potential curves, ranging in likelihood based on the observed data. As more simulations are run, the surrogate model (in essence our map of the parameter space) is updated with new points. Areas where the data is denser will have lower uncertainty between points, where regions with fewer points will possess more uncertainty. The Gaussian process balances searching these two regions through exploration versus exploitation. Meaning when searching for a minimum if it starts to find a promising area in the parameter space it will more finely explore that area, while simultaneously in other runs test unknown areas where data points are fewer and possess the potential for unrealized optimums. 

\begin{figure}
  \centering
  \vspace*{\fill}
  \includegraphics[width=1\linewidth]{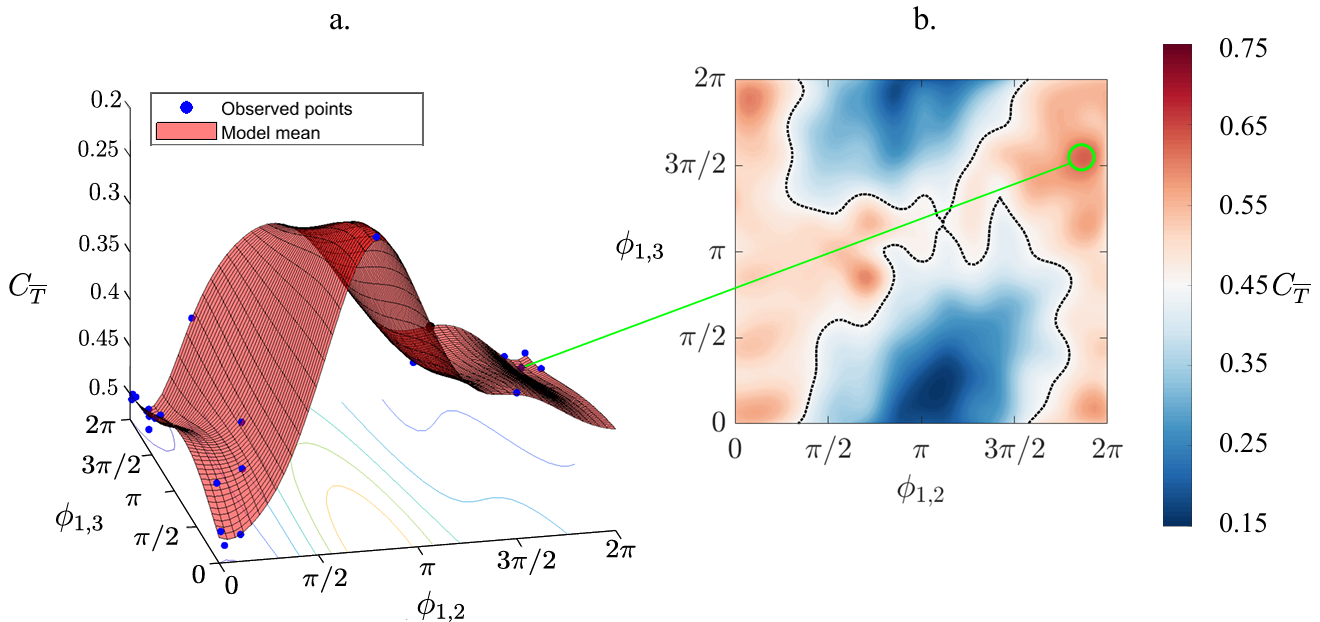}
  \vspace*{\fill}
  \caption{Bayesian optimization model of the 3-fin system for thrust as it varies with fin phases (a) in comparison to the completed parameter space (b).}
  \label{fig:bayesian_explanation}
\end{figure}

Before moving to a completely unknown parameter space, we first tested the Bayesian optimizer against the known three-fin cases explored in Sec.~\ref{sec:3fin}. For this validation case, the optimizer was initialized at \(\phi_{1,2}=\phi_{1,3}=0\) and allowed a maximum of 20 objective evaluations. The optimization was implemented using MATLAB's built-in \texttt{bayesopt} routine with the \texttt{expected-improvement-plus} acquisition function, and was performed serially rather than in parallel. Additionally, since the optimization framework is formulated as a minimizer, the objective supplied to the optimizer was the negative of thrust, such that minimizing the objective corresponds to maximizing thrust. Due to limited computing resources, we did not run multiple restart campaigns in this implementation; therefore, we recognize that although in this limited framework we find the best solution, it may not be the absolute global optimum. The computer hardware used for the Bayesian optimization runs is outlined in Appendix B.

In figure \ref{fig:bayesian_explanation}, we show the optimizer maximizing thrust for the three-fin system using fin phase as the design input. Each tested point is shown within the bounded design space, while the surface represents the Gaussian-process surrogate model used to estimate performance between sampled points. The surrogate captures the dominant low-performance regions, such as the broad area near \(\phi_{1,2}=\pi\), as well as the high-performance region near \(\phi_{1,2}=0\). The maximum-thrust configuration identified by the optimizer was also consistent with a maximum observed in the complete parameter dataset.

To assess search efficiency, we compare this result to the full performance map in figure ~\ref{fig:bayesian_explanation}, which required 225 simulations to evaluate the parameter space exhaustively. In contrast, the Bayesian optimization identified the same high-performing phase region after only 20 CFD evaluations. In the present implementation, the stopping criterion was the maximum evaluation budget, and the Gaussian-process surrogate hyper-parameters were handled internally by MATLAB's default \texttt{bayesopt} implementation rather than prescribed manually.

\begin{figure}
  \centering
  \vspace*{\fill}
  \includegraphics[width=1\linewidth]{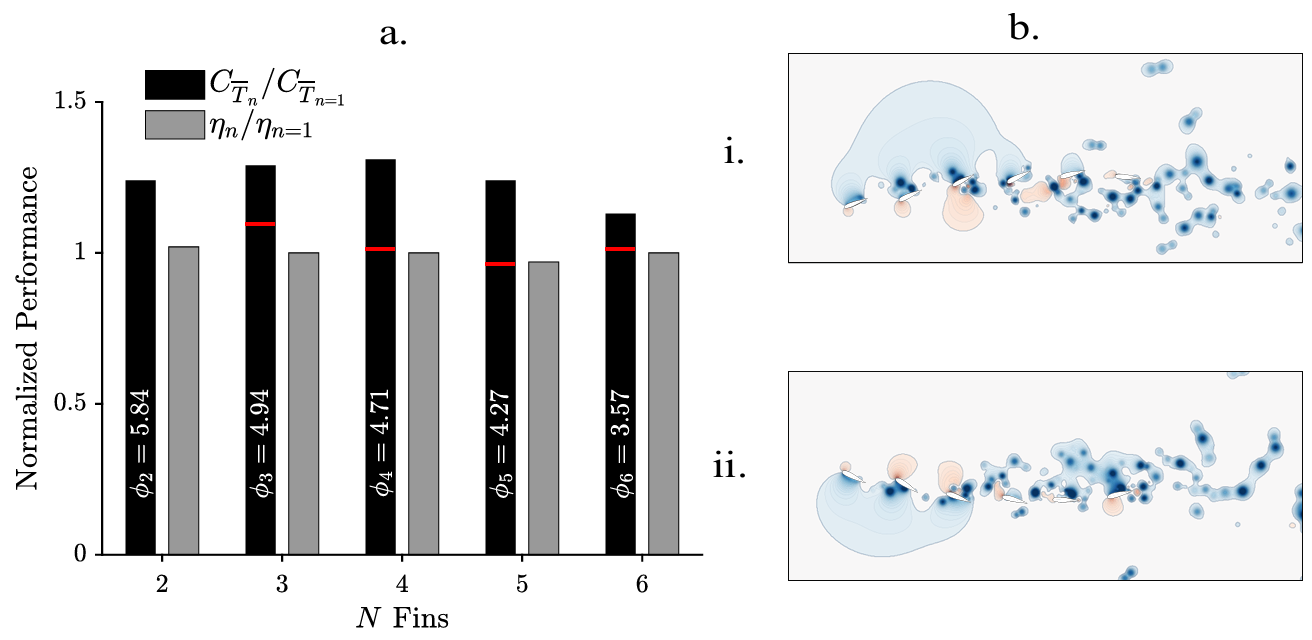}
  \vspace*{\fill}
  \caption{Thrust and efficiency of $N$-fin systems optimized for thrust with reference lines representing the performance of a system having equal phase offsets; with reference flow fields of the 6-fin case at $t/T$ = 0.5 (b.i) and 1 (b.ii). A supplemental file has an animation of the $N=6$ flow field.}
  \label{fig:bayesian_truncated_thrust}
\end{figure}

To expand beyond the 3-fin system, we do not need to optimize across the entire parameter space---instead we can make use of what we learned from our detailed analysis surrounding figure \ref{fig:3fin_thrust_varied_spacing}. Unless the fins are too close together ($S\leq 0.5$), downstream fins do not impact the performance of upstream fins. Thus, when exploring a system with an additional fin, we need only to optimize the kinematics of that specific additional fin, i.e., sequential optimization. In figure \ref{fig:bayesian_truncated_thrust} we see the optimal results the $N=2-6$ fin systems, maximizing thrust. The data has been normalized by the performance of a single fin, allowing us to see the relative increase in system performance by the addition of each fin. Here we also show the efficiency of the system that is thrust-optimized. With this optimization, we found that the addition of fins beyond a single fin system show a slight increase in performance up to $N=4$ fins, though past this point there are diminishing returns. This is likely because, as more fins are added the wake of the system becomes increasingly chaotic. Thus, it becomes increasingly difficult for additional fins to take advantage of the wake vortex structure.   

Interestingly, we see the optimal phase of the additional fins to be unequal, where one might logically assume that the best phase between fin 1-2 is the same as the best phase between 4-5, with spacing and speed being constant. This is likely because of the chaotic wake and existence of multiple vortex structures for the additional fins to interact with. To be sure the optimizer was finding ideal performance, on this plot we compare to the thrust of a system where every additional fin is added at the same phase offset that matched the ideal phase of the second fin (i.e., $\phi_{1,2} = \phi_{2,3} = \phi_{3,4} = \phi_{4,5} = \phi_{5,6} = 5.84$). The performance of this case is shown as red markers on the thrust levels. It is clear that a decaying phase difference between subsequent fins outperforms what would be a logical first guess at the ideal system without optimization.

\begin{figure}
  \centering
  \vspace*{\fill}
  \includegraphics[width=1\linewidth]{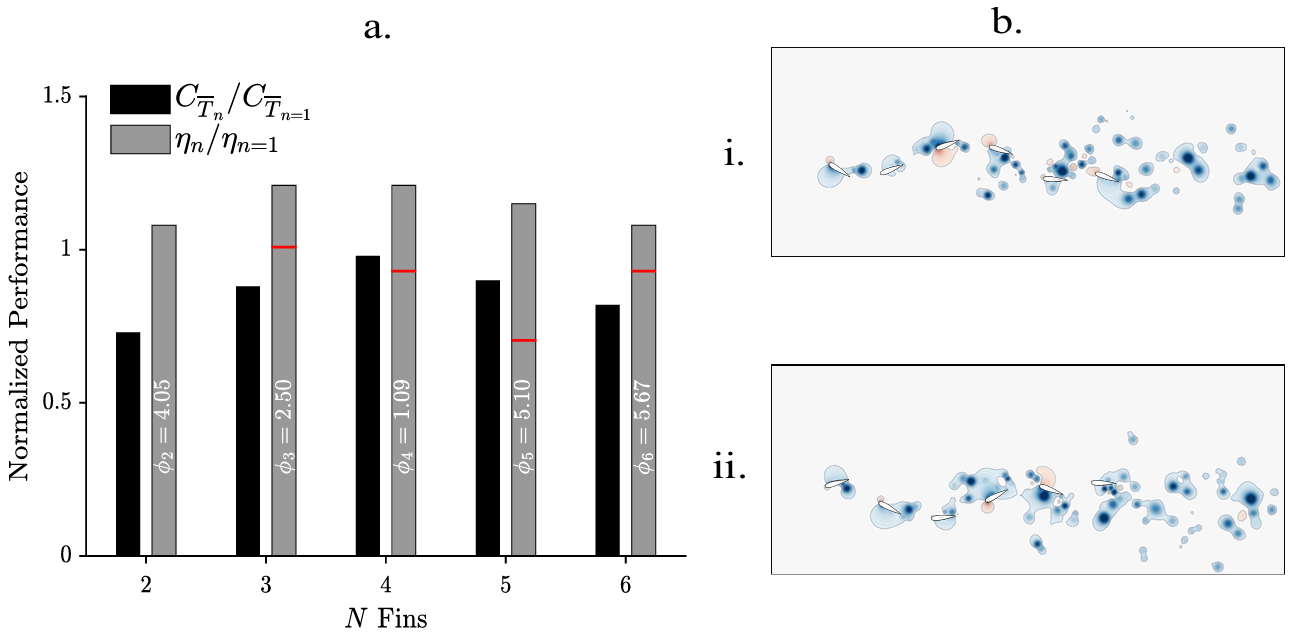}
  \vspace*{\fill}
  \caption{Thrust and efficiency of $N$-fin systems optimized for efficiency with reference lines representing the performance of a system having equal phase offsets; with reference flow fields of the 6-fin case at $t/T$ = 0.5 (b.i) and 1 (b.ii). A supplemental file has an animation of the $N=6$ flow field \cite{SM}.}
  \label{fig:bayesian_truncated_eta}
\end{figure} 

In figure \ref{fig:bayesian_truncated_eta} we do the same optimizing scheme, but this time to maximize efficiency. The optimal efficiency relative to the single fin case and corresponding thrust are shown. As with the thrust, we see a climb in the optimal efficiency with fin number then decay. The peak potential efficiency for the system occurs at $N=3-4$ fin systems operating about 20\% more efficiently than a single fin. The thrust penalty when operating at optimal efficiency is non-negligible, though it is notable that the $N=4$ case has relatively high efficiency with almost no thrust losses when compared to a single fin.

In both figures \ref{fig:bayesian_truncated_thrust} and \ref{fig:bayesian_truncated_eta} we see a snapshot of the flow field for the optimal $N=6$ cases. It is quite clear that the dynamics of the fin systems have coherence---each appears to make a traveling wave downstream. (This is most evident by observing the animations in the supplemental files that correspond to these figures \cite{SM}). To maximize thrust, the fin system traveling wave is relatively long in wavelength, while maximizing efficiency yields a shorter wavelength in comparison.

\section{Conclusion}

We investigated the performance capabilities of multi-fin inline oscillatory propulsion systems using an immersed boundary method flow solver. Of particular interest were the thrust and propulsive efficiency of the system and how they varied with number of fins, fin spacing, and phase of the fins. For the majority of the work the Strouhal number was fixed at $St = 0.32$, representing a common swimming condition. 

First, we validated our flow solver by exploring a 2-fin system and compared our results to those from literature. Our results matched nicely for replicated conditions. In multi-fin systems, the performance of the downstream fins rely on timing in the wake governed by the phase difference between the fins, the spacing, and the freestream velocity. We showed that the phase modifier $\phi_0 \sim 2 \pi S f / U_\infty$ captured this relationship and collapsed the performance results of a 2-fin system  (and then subsequently confirmed for a 3-fin system) across a wide range of these parameters.

Then we parametrically explored the 3-fin system varying the fin phases and spacing from $S=0.5-10$. As with the 2-fin system, the timing is critical---thrust and efficiency could vary between $C_{\overline{T}}=0.15-0.6$ and $\eta=0.2-0.45$ depending on $\phi_0$. In exploring the time-history of the forces, boosts in thrust occurred when the shed vortices from the upstream fins passed near the suction side of the downstream fin. Conversely, the shed vortex colliding with the foil or passing near the underside is detrimental to performance.  Overall, we found that the performance of a downstream fin was governed almost entirely by the nearest upstream fin. Only when fins were spaced very close together ($S=0.5$) would an upstream fin feel any meaningful impact of the downstream fin in the wake.  

A major implication of the detailed 3-fin study was that if you have an optimized $N$-fin system, adding an additional fin downstream did not change the optimal configuration of the upstream fins. Thus, optimizing larger fin systems can be done sequentially for every new fin instead of considering the dynamics of the whole fin set for each new optimization. We conducted performance optimizations for up to $N=6$ fin systems, maximizing either thrust or efficiency of the system. These systems could reliably perform with up to 30\% more thrust per fin or 20\% more efficiency when compared to a single fin. Generally, performance of the system peaked at $N=4$ fins, then gradually decayed. The diminishing performance gains with increasing fin number appear to be associated with progressively complex interactions among vortical structures shed by upstream fins. These interactions make it increasingly difficult for downstream fins to maintain favorable vortex encounters, although further wake characterization is needed to establish the dominant mechanism.

Finally, it is important we consider a few limitations of this work. First, the Strouhal number is not varied explicitly throughout the study, despite being one of the main kinematic variables in swimming systems. Based on past results \cite{Muscutt2017} and the physics uncovered in this work (primarily in the new parameter $\phi_0$), we do not anticipate that changing the Strouhal number would alter these conclusions provided that the wake is still coherent and vortex dominated. However, other parameters like three-dimensionality and turbulence likely have an impact, as they are what govern the large-scale vortex dynamics and breakdown in the wake. If these cause vortex breakdown more quickly, it could be that the performance benefits from the fin-fin interaction would diminish. Conversely, it may be that more rapid breakdown of the structures from the furthest upstream fins would lead to a less chaotic wake for a downstream fin, with the dominant structure coming from the nearest upstream neighbor. It may be that the enhanced vortex breakdown leads to less diminishing returns than found in this work. We look forward to considering these parameters in future studies. 

\section*{Acknowledgments}
This work was completed with the funding from the Office of Secretary of Defense under the management of Dr. Bob Brizzolara. We would also like to thank Dr. Gabriel Weymouth for their advisement on WaterLily code implementation and Dr. Frank Tricouros for paper editing and guidance.
\section*{Funding}
The work was funded under an OSD STTR grant number W911NF23C0053.

\section*{Author contributions}
PTG led the research, data acquisition, data analysis, and wrote the article. EJW programmed, tested, and validated the simulation code. TVB oversaw the research direction, secured funding, assisted in data acquisition and analysis, and edited the article.

\bibliographystyle{unsrt}
\bibliography{apssamp}

\clearpage
\appendix
\section{Convergence Study}
\label{app:convergence}

\subsection{Kinematic convergence}

We begin with a kinematic refinement study using a baseline configuration of three in-line fins spaced two chord lengths apart and oscillating in phase (i.e., no phase offset between fins). The computational domain extends $30c$ in the streamwise direction and $25c$ in the transverse direction. Each fin has a NACA 0020 profile and undergoes a prescribed sinusoidal heave of amplitude $h_0 = c$ combined with pitching motion of $\pm 30^\circ$. Simulations were conducted for 10 oscillation cycles, with 100 data points recorded per cycle.

\begin{figure}[t]
\centering
\includegraphics[width=0.9\textwidth]{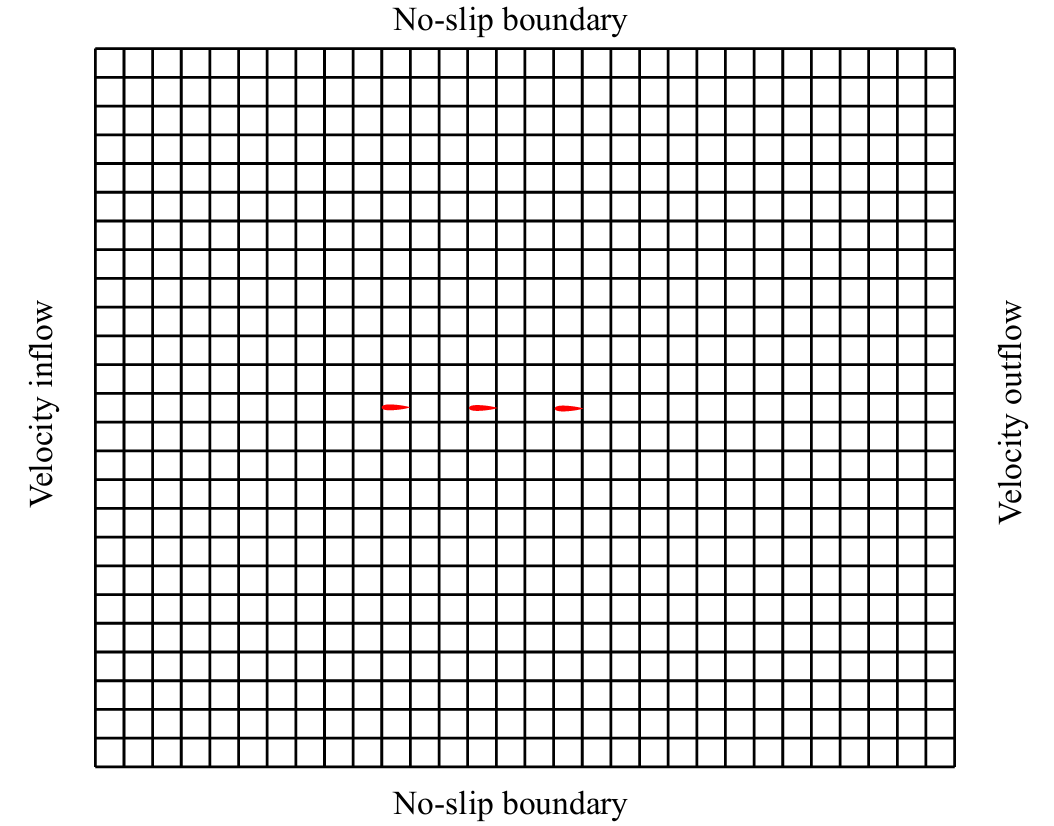}
\caption{Computational domain schematic for the $N=3$ fin system. The domain extends $30c$ in the streamwise direction and $25c$ in the transverse direction. Grid resolution corresponds to either 32 (coarse) or 64 (fine) cells per chord.}
\label{fig:cdomain}
\end{figure}

\begin{figure}[t]
\centering
\includegraphics[width=\textwidth]{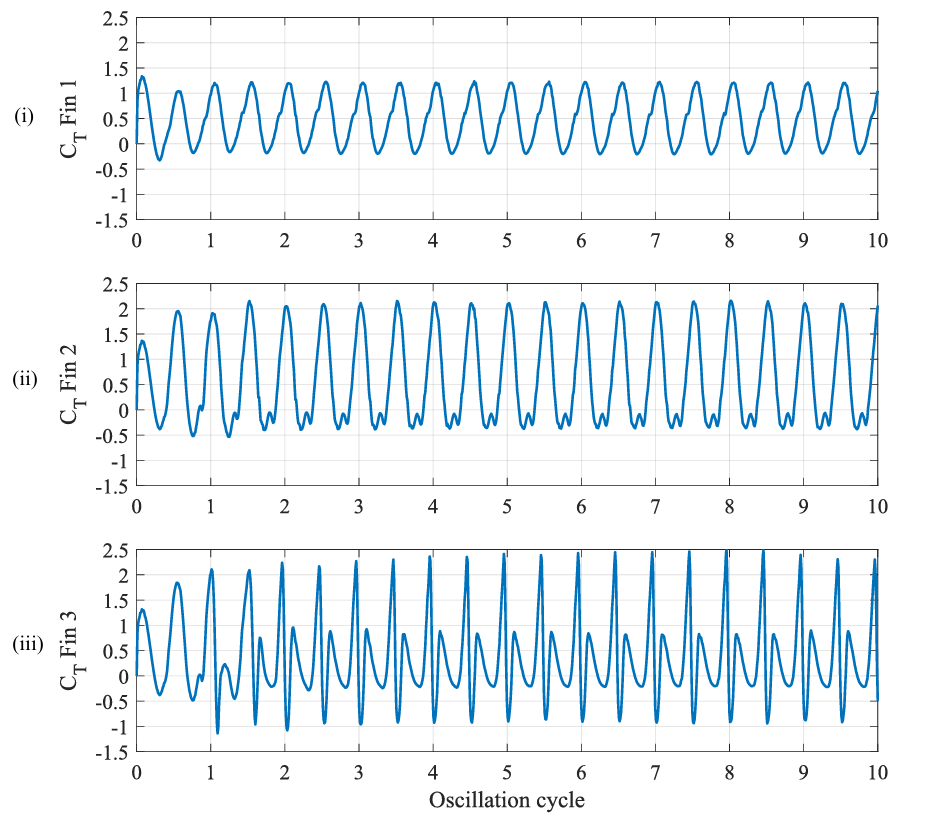}
\caption{Instantaneous coefficient of thrust, $C_T$, for each fin in a three-fin system over 10 oscillation cycles.}
\label{fig:cycle}
\end{figure}

A schematic of the computational domain is shown in Fig.~\ref{fig:cdomain}. The instantaneous coefficient of thrust, $C_T$, for each fin is plotted over all cycles in Fig.~\ref{fig:cycle}. To assess convergence, cycle-averaged thrust values were computed for the third cycle (C3), the final cycle (C10), and the average of the final five cycles (L5), as summarized in Table~\ref{tab:cycle}.

\begin{table}[t]
\caption{Relative percent difference in mean thrust coefficient, $\overline{C_T}$, computed using different cycle averaging windows.}
\label{tab:cycle}
\centering
\small
\begin{tabular}{lccccc}
\toprule
 & $\overline{C_T}$ (C3) & $\overline{C_T}$ (C10) & $\overline{C_T}$ (L5) & \% Diff (C3 vs.\ C10) & \% Diff (C3 vs.\ L5) \\
\midrule
Fin 1 & 0.477 & 0.474 & 0.472 & 0.63\% & 1.06\% \\
Fin 2 & 0.622 & 0.622 & 0.626 & 0.00\% & 0.64\% \\
Fin 3 & 0.315 & 0.320 & 0.333 & 1.56\% & 5.41\% \\
\bottomrule
\end{tabular}
\end{table}

The average thrust coefficients for fins 1 and 2 show less than $1\%$ deviation between the third and final cycles. Fin 3 exhibits slightly larger variation, with a $1.56\%$ difference relative to the final cycle and $5.41\%$ relative to the average of the last five cycles. Based on these results, the third oscillation cycle is deemed sufficient for estimating system performance in coarse and fine resolution studies. For cases with $N \geq 3$ fins, additional cycles are added to ensure the vortices reach the added downstream structure. For $N \geq 3$ fins, the $N$-fin system will run at $N$ cycles and the performance will be taken from the final cycle.

\subsection{Temporal convergence}

A temporal resolution study was conducted to evaluate the number of data points required per oscillation cycle. Baseline simulations recorded 100 points per cycle, while reduced-output simulations stored 33 points per cycle (corresponding to 99 points over three cycles). Importantly, the solver time step remained unchanged; only the frequency of data output was reduced.

\begin{figure}
\centering
\includegraphics[width=0.75\textwidth]{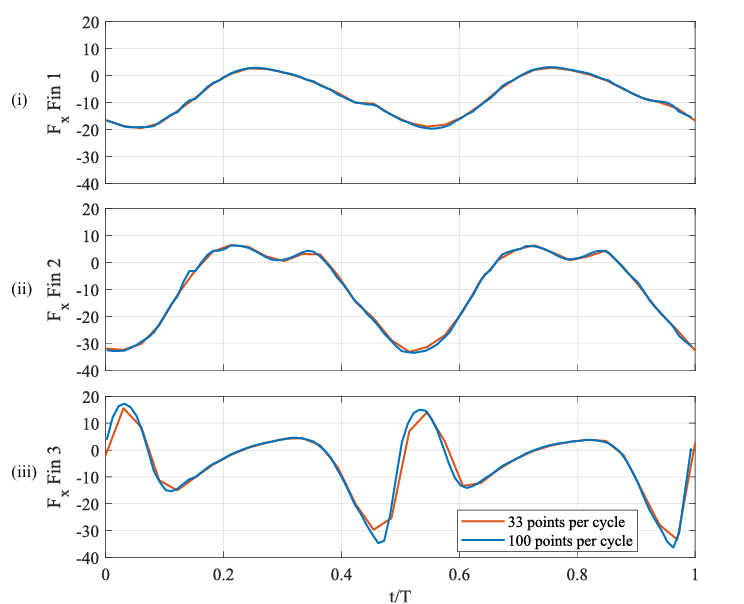}
\caption{Instantaneous streamwise force for each fin over one oscillation cycle, comparing 33 and 100 saved points per cycle.}
\label{fig:temporal}
\end{figure}

Figure~\ref{fig:temporal} compares the instantaneous streamwise force for each fin over a representative oscillation cycle. The waveform structure, including peak magnitudes and phase alignment, is well preserved between the two sampling strategies. Minor discrepancies are observed primarily in the third fin, which is more sensitive to upstream wake interactions.

Cycle-averaged thrust values are compared in Table~\ref{tab:temporal}. The largest deviation occurs for fin 2 ($4.61\%$), while all other differences are smaller. Despite these differences, the relative performance trends and ranking between fins remain unchanged.

\begin{table}[t]
\caption{Temporal resolution comparison between 33 and 100 saved points per cycle.}
\label{tab:temporal}
\centering
\small
\begin{tabular}{lccc}
\toprule
 & $\overline{C_T}$ (33) & $\overline{C_T}$ (100) & \% diff \\
\midrule
Fin 1 & 0.48 & 0.49 & 2.04 \\
Fin 2 & 0.62 & 0.65 & 4.61 \\
Fin 3 & 0.32 & 0.32 & 0.00 \\
System Avg & 0.47 & 0.49 & 4.08 \\
\bottomrule
\end{tabular}
\end{table}

These results indicate that reduced temporal sampling preserves the essential dynamics of the force signal while significantly reducing storage and post-processing costs. Therefore, 33 points per cycle were used for coarse parameter sweeps, while 100 points per cycle were retained for high-fidelity simulations.

\subsection{Mesh convergence}

A grid cell mesh refinement study was performed to assess grid resolution requirements. A reference simulation using 128 cells per chord was first conducted based on the validation conducted in a previous study using this flow solver \cite{Muscutt2017}. At $Re = 5000$, the laminar boundary layer thickness scales as $\delta \sim 5c/\sqrt{Re} \approx 0.07c$, corresponding to approximately nine grid cells at this resolution. This was deemed as sufficient to resolve near-wall gradients.

\begin{figure}
\centering
\includegraphics[width=0.6\textwidth]{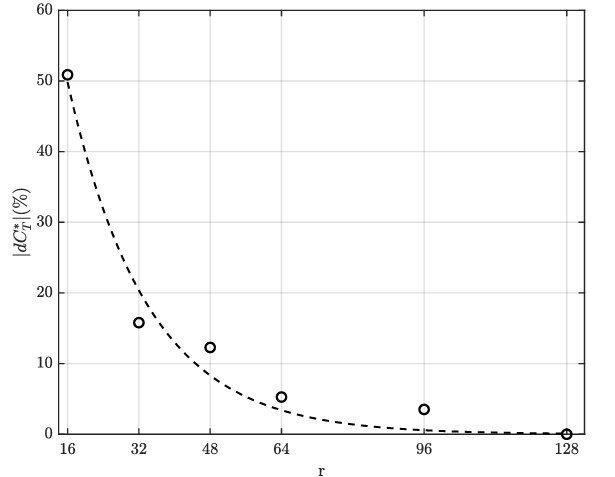}
\caption{Grid refinement study showing relative error in system-average $\overline{C_T}$ versus cells per chord, using 128 cells per chord as the reference.}
\label{fig:grid}
\end{figure}

We then conducted a refinement study in which cell count per fin was systematically reduced, and the resulting percent difference in average coefficient of thrust between each reduced run and the 128 cell run was plotted, results shown in figure \ref{fig:grid}.

At 32 cells per chord, the predicted system-average thrust coefficient was approximately $16\%$ lower than the reference solution. Increasing the resolution to 64 cells per chord reduced this deviation to approximately $5.26\%$, as shown in Fig.~\ref{fig:grid}. Based on this trade-off,  32 cells per chord were employed for large parameter sweeps, to highlight regions of interest in vortex and fin interactions. Then with specific cases of interest, 64 cells per chord would be used for high-fidelity runs, where we want to directly know the performance of a specific system. 

\begin{figure}[t]
\centering
\includegraphics[width=0.65\textwidth]{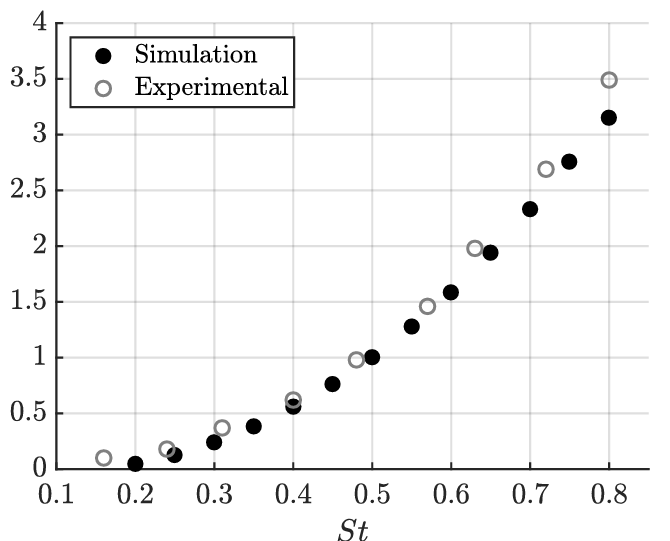}
\caption{Cycle-averaged thrust coefficient, $\overline{C_T}$, as a function of Strouhal number for a single fin.}
\label{fig:ch1_validation}
\end{figure}

To assess whether the coarse grid captures relevant physics, simulations were compared with experimental results from Van Buren \textit{et al.}~\cite{VanBuren2019}. As shown in Fig.~\ref{fig:ch1_validation}, the simulations reproduce the expected increase in $\overline{C_T}$ with Strouhal number and capture the correct performance trends. The differences between the experiment and simulation can be attributed to the Reynolds number differences, which have slight impact on performance \cite{senturk2019reynolds}. 

\begin{figure}[t]
\centering
\includegraphics[width=1\textwidth]{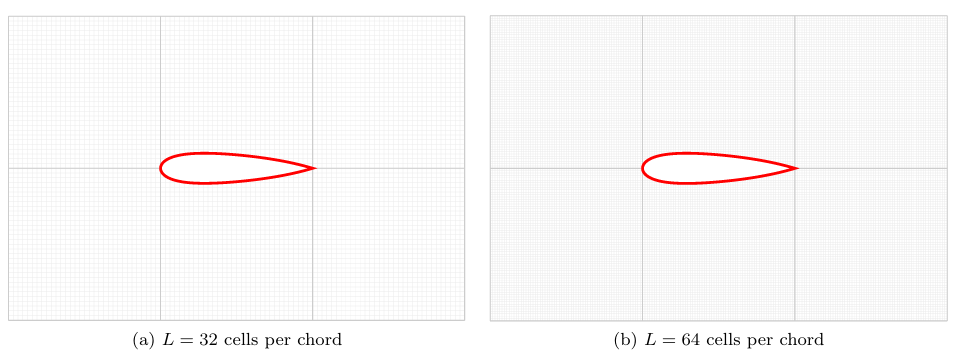}
\caption{Comparison of computational mesh resolution near the foil.}
\label{fig:mesh_compare}
\end{figure}

Finally, Fig.~\ref{fig:mesh_compare} compares the computational mesh near the foil for $L=32$ and $L=64$ cells per chord. The finer grid better resolves the immersed boundary geometry, while the coarser grid enables faster simulations for large-scale parameter studies.

\newpage
\section{Computational Hardware for Optimization Runs}
\label{app:hardware}

To support reproducibility of the Bayesian optimization results presented in Section~3.3, we document the computational hardware and execution strategy used for the CFD evaluations comprising the objective function. Each evaluation corresponds to a full simulation of oscillating hydrofoils using a two-dimensional viscous incompressible flow solver (WaterLily, implemented in Julia), and thus the total optimization cost is strongly dependent on hardware configuration and parallelization strategy.

All optimization runs were executed on a dual-socket workstation equipped with:
\begin{itemize}
    \item CPU: 2 $\times$ Intel Xeon Silver 4215R processors
    \item Total physical cores: 16 (32 logical processors)
    \item Base clock frequency: 3.19~GHz
    \item Sustained frequency under load: approximately 2.0--3.4~GHz
    \item Memory: 128~GB DDR4 (2400~MHz)
    \item Storage: NVMe solid-state drive (approximately 1.9~TB capacity)
\end{itemize}

Simulations were performed using CPU-based parallelism, with the number of threads per simulation explicitly controlled to avoid oversubscription of memory bandwidth and to maintain stable clock frequencies during long-duration runs. Although 32 logical processors are available, it was observed that using all available threads degraded performance due to cache contention and synchronization overhead associated with stencil-based CFD operations and Poisson solves. Accordingly, 16 threads was found to be optimal.
\end{document}